\documentclass[conference]{IEEEtran}
\IEEEoverridecommandlockouts
\usepackage{cite}
\usepackage{amsmath,amssymb,amsfonts}
\usepackage{algorithmic}
\usepackage{textcomp}
\usepackage{subcaption}
\usepackage{tcolorbox}
\usepackage{comment}
\usepackage{tabularx}     
\usepackage{graphicx}
\usepackage[flushleft]{threeparttable} % for table footnotes
\usepackage{xcolor}
\usepackage{hyperref}
\def\BibTeX{{\rm B\kern-.05em{\sc i\kern-.025em b}\kern-.08em
    T\kern-.1667em\lower.7ex\hbox{E}\kern-.125emX}}
\begin{document}

\title{Comparing ML-Specific and General Python Code Smells Across Project 
Characteristics\\
%\thanks{}
}

\author{
\begin{minipage}[t]{0.32\linewidth}
\centering
\textbf{Halimeh Agh}\\
Institute of Software Engineering\\
University of Stuttgart\\
Stuttgart, Germany\\
halimeh.agh@iste.uni-stuttgart.de
\end{minipage}%
\hfill
\begin{minipage}[t]{0.32\linewidth}
\centering
\textbf{Betül Cimendag}\\
TUM School of Computation, Information and Technology\\
Technical University of Munich\\
Munich, Germany\\
betul.cimendag@tum.de
\end{minipage}%
\hfill
\begin{minipage}[t]{0.32\linewidth}
\centering
\textbf{Stefan Wagner}\\
TUM School of Computation, Information and Technology\\
Technical University of Munich\\
Heilbronn, Germany\\
stefan.wagner@tum.de
\end{minipage}
}

%\and
%\and
%\and

\maketitle 

\begin{abstract}
Machine learning systems consist of general-purpose code as well as machine-learning-specific code. While ML-specific code smells have been identified, their connection to project characteristics and their interaction with overall code quality are not well understood. Without this knowledge, quality assurance strategies remain one-size-fits-all, failing to account for the contextual factors that drive technical debt in ML systems. We present empirical evidence by examining how six project features (size, age, contributors, commit frequency, CI/CD adoption, and domain) relate to both ML-specific and general Python code quality in 279 open-source ML projects on GitHub. Using CodeSmile for ML code smells and Pylint for general Python smells, our results show: (1) ML code smells are 41--94 times less frequent than general Python smells; (2) commit frequency and domain are significantly associated with ML-specific quality, while project size, team size, age, and CI/CD adoption are not, challenging traditional views on technical debt; (3) general Python smells are not linked to any project characteristic, indicating systemic coding issues that are independent of project context; (4) domains that suffer most from ML-specific smells are not necessarily the same domains that suffer most from general Python smells, necessitating tailored quality strategies for each smell type. MLOps often involves configuration issues, Reinforcement Learning faces challenges with tensor manipulation, and Computer Vision encounters problems with GPU workflows. Overall, ML code quality depends on domain-specific practices and specialized CI/CD quality gates, as standard automation often overlooks domain-specific correctness problems.
\end{abstract}


\section{Introduction}
Machine Learning (ML) has become increasingly prevalent in modern software systems, with applications including autonomous vehicles, natural language processing (NLP), medical diagnosis, and fraud detection in financial transactions \cite{b1},\cite{b2}. As ML-enabled systems grow in complexity and scale, ensuring their quality has become a significant challenge for researchers and practitioners \cite{b3},\cite{b4}. However, these systems are particularly prone to accumulating technical debt due to their experimental nature and intricate infrastructure requirements \cite{b5}. One aspect of technical debt is the presence of code smells, which are indicators of poor design and implementation choices that can degrade software quality over time \cite{b6},\cite{b7}. Code smells can include potential defects in the source code, opportunities for refactoring, and violations of coding standards \cite{b8}. Research has indicated that code smells primarily impact maintainability, understandability, and overall complexity. Furthermore, early detection of these issues can significantly reduce maintenance costs \cite{b9},\cite{b10}.

While code smells have been extensively studied in traditional software systems \cite{b7},\cite{b9}, their presence in ML systems introduces unique challenges. These challenges arise from distinct characteristics of ML, including a heavy reliance on mathematical operations, the integration of multiple specialized libraries, iterative experimentation, and complex data preprocessing pipelines \cite{b8}. Recent research has started to examine these ML-specific quality issues from various angles. For example, Van Oort et al. \cite{b8} analyzed the prevalence of traditional code smells in ML projects using Pylint, revealing widespread issues related to code duplication and dependency management. Additionally, Zhang et al. \cite{b11} introduced a catalog of 22 ML-specific code smells (ML-CSs) that identify implementation anti-patterns unique to ML pipelines. These include problems such as improper handling of random seeds, inefficient data operations, and incorrect API usage, all of which can lead to degraded performance, increased error rates, and compromised reproducibility.

Despite these advances, a critical gap remains in our understanding of how ML-CSs relate to general Python code smells in real-world ML projects. While existing studies have examined either ML-CSs \cite{b11},\cite{b12} or general code quality issues \cite{b8} in isolation, no comprehensive investigation has systematically compared the prevalence and characteristics of both smell types within the same set of ML projects. More importantly, it remains unclear how these two types of smells differ in their correlation with key project attributes (e.g., project size), whether traditional quality practices (such as CI/CD and long-term maintenance) impact both types in the same way, and which domains encounter challenges with each type of smell. Understanding these differential relationships is crucial for developing effective quality assurance strategies that can prioritize the most significant issues based on the project context. For example, knowing that ML-CSs are driven by domain and development activity rather than project size or CI/CD adoption allows teams to complement general automated linting with domain-tailored, ML-specific quality checks that target the actual drivers 
of ML technical debt.

To address this gap, we present a large-scale empirical study to understand how ML-specific and general Python code smells relate differently to key project characteristics, and investigate this relationship across 279 open-source ML projects from the NICHE dataset \cite{b13}. We employ two complementary detection tools: CodeSmile \cite{b12}, a static analysis tool specifically designed to detect 12 ML-CSs, and Pylint \cite{b8}, a widely used general-purpose Python linter. Our analysis encompasses 132,067 Python files, containing over 17,911,446 lines of code. It systematically examines how both types of smells correlate with project characteristics, including size, age, domain, CI/CD adoption, and the number of contributors.

Specifically, we address the following research questions:
\noindent\textbf{RQ1:} How do the densities of ML-CSs and general Python code smells compare across ML projects?

\noindent\textbf{RQ2:} What are the relationships between ML-CSs and project characteristics?

\noindent\textbf{RQ3:} What are the relationships between general Python code smells and the same project characteristics?

\noindent\textbf{RQ4:} What are the differences in how ML-CSs and general Python code smells correlate with project characteristics?

The main contributions of this paper are:
\begin{itemize}
  \setlength\itemsep{0pt} % remove extra space between items
  \setlength\itemindent{1em} % small horizontal indent
  \item A large-scale comparative study of ML-specific and general Python code smells across 279 real-world ML projects
  \item An updated, enriched NICHE dataset with additional project features
  \item Empirical evidence showing  correlation patterns between smell types and project features
  \item Evidence-based recommendations for ML practitioners on prioritizing quality assurance efforts based on project context
\end{itemize}
To ensure reproducibility and enable future research, we provide a comprehensive replication package\footnote{\url{https://github.com/HalimehAgh/ml-code-smells-replication}} containing: (i) the extended NICHE dataset with all newly extracted features, (ii) the CodeSmile and Pylint analysis outputs for all 279 projects, (iii) statistical analysis scripts and intermediate results.

\section{Related Work}

%\subsection{Maintaining the Integrity of the Specifications}

Research on code quality in ML systems has progressed along two main paths: studies exploring general Python code quality issues in ML projects and investigations into ML-CSs. We review both areas, emphasizing their findings and limitations that drive our comparative study.

\textbf{General Python Code Quality in ML Projects.} Several studies have examined the prevalence of general Python code smells in ML and data science projects. Van Oort et al. \cite{b8} analyzed 74 ML projects using Pylint, finding that code duplication is common and dependency management issues pose major threats to reproducibility and maintainability. Their qualitative analysis showed that duplicate code mainly arises from multiple experimental model variations, while dependency specification problems hinder project setup and replication. Simmons et al. \cite{b14} carried out a large-scale comparison of 1048 data science projects against 1099 non-data science projects, discovering that data science projects have significantly higher rates of functions with excessive parameters and local variables. They suggested that traditional software engineering conventions may be inappropriate for data science code due to its mathematical nature and domain-specific needs. However, both studies focused on prevalence patterns without exploring how code smells relate to project characteristics such as size, age, or development practices.

Recently, researchers have started exploring the relationships between software engineering practices and code quality in ML projects. Giordano et al. \cite{b15} conducted a longitudinal study on the spread of code smells in ML-enabled systems, examining the activities that lead developers to introduce smells and how long these smells persist as the project progresses. Their results suggested that smell variation does not follow specific time patterns and that smells can persist for several years. Building on this research, Giordano et al. \cite{b16} conducted a study of 566 Python ML projects from the NICHE dataset, examining relationships between eight software engineering practices and ten Python-specific code smells. They found that projects implementing SE practices show significantly fewer code smells, with Continuous Integration being negatively associated with the Complex Container Comprehension smell. Tang et al. \cite{b17} studied technical debt in 26 ML projects, identifying various code anti-patterns and emphasizing the high frequency of duplicated code, while Cardozo et al. \cite{b18} analyzed 24 reinforcement learning (RL) projects, verifying that traditional code smells are more common in ML systems than in traditional software. Despite these advances, existing studies on general Python code quality in ML projects have limitations: some focus only on prevalence without examining relationships with project characteristics \cite{b14},\cite{b8},\cite{b17},\cite{b18}, others examine temporal changes instead of project-level factors \cite{b15}, while recent correlation studies analyze only general Python smells without considering ML-specific quality issues \cite{b16}.

\textbf{ML-Specific Code Smells.} Recognizing that ML systems are particularly prone to technical debt \cite{b5} and that traditional code smells alone are insufficient for capturing quality issues specific to ML development, Zhang et al. \cite{b11} introduced a catalog of 22 ML-CSs based on empirical analysis of white and grey literature. Building on this foundation, several researchers have developed automated methods to detect ML-CSs. Recupito et al. \cite{b12} developed CodeSmile, a static analysis tool for identifying ML-CSs, and conducted a large-scale lifecycle study analyzing over 400,000 commits from 337 NICHE projects. Their research examined the prevalence, introduction, removal, and persistence of ML-CSs, finding that these smells are often introduced during file modifications for new feature development and usually eliminated during enhancement or refactoring. Notably, Recupito et al. also explored the relationship between ML-CS prevalence and project characteristics, using statistical tests to compare project sizes (small, medium, large) and CI adoption. Recent work has further improved ML-specific smell detection capabilities. Hamfelt et al. \cite{b19} developed MLpylint, a static analysis tool that detects 20 ML-CSs using Abstract Syntax Tree analysis, validating it on 160 open-source projects with feedback from ML experts. Mahmoudi et al. \cite{b20} introduced SpecDetect4AI, which combines a Domain-Specific Language for smell specification with an extensible static analysis tool, and tested it on 826 AI-based systems. However, these studies focus on detecting and understanding ML-CSs and do not explore how they compare with general Python code quality patterns or how different smell types relate to project characteristics.

%While previous research has focused on either ML-specific smells [11, 12, 18, 19] or general Python code quality [6, 8, 14-17] separately, no study has systematically compared both types of smells within the same projects to explore how they relate to project features. It is crucial to determine if these smells react similarly to factors like CI/CD adoption or commit frequency, and whether issues in one smell type also affect the other. Our study addresses this gap by conducting a large-scale comparative analysis of 279 ML projects.

\section{Study Design}
\begin{figure}[t]
  \centering
  \includegraphics[width=1.05\linewidth]{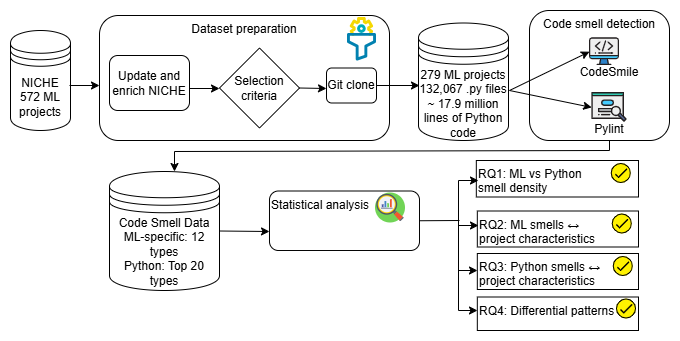}
  \caption{Overview of the research methodology.}
  \label{fig:fig1}
\end{figure}

Our main goal is to understand how ML-specific and general Python code smells relate differently to key project characteristics. To reach this goal, we formulated four research questions: RQ1 compares the densities of ML-specific and general Python code smells across ML projects; RQ2 examines the relationships between ML-CSs and project characteristics, including size, age, commit frequencies, domain, CI/CD adoption, and number of contributors; RQ3 investigates the connections between general Python code smells and the same project features; and RQ4 analyzes how these two types of smells differ in their correlations with project characteristics. Figure~\ref{fig:fig1} illustrates our research methodology, showing the flow from dataset selection through code smell detection and statistical analysis to final results. Each phase is explained in detail in the following subsections.

\subsection{Dataset Description and Project Selection}\label{AA}
To enable empirical studies of ML code quality, Widyasari et al. \cite{b13} introduced NICHE, a curated dataset of 572 Python machine learning projects. The dataset was created by selecting projects that demonstrate evidence of strong software engineering practices across eight dimensions: architecture, community, continuous integration, documentation, history, issues, license, and unit testing. Each project is classified as "Engineered" or "Non-Engineered" based on these quality dimensions. The dataset provides project metadata, including GitHub repository information, star counts, commit numbers, and lines of code. Since the NICHE repositories were cloned in 2020 and repositories evolve continuously, we updated several metrics to reflect the current project states. Specifically, we refreshed four existing NICHE metrics: star counts, commit numbers, lines of code (LOC), and last commit date. Additionally, we extracted two features necessary for our project selection criteria: the number of contributors and the number of branches. The update process accessed 565 of 572 repositories (98.8\%), with 7 repositories inaccessible due to deletion or privatization since 2020. Technical implementation details, including API query specifications and error handling procedures, are available in our replication package.
To focus on active, collaborative projects, we applied five filtering criteria sequentially: recent commit activity (2024 or later) \cite{b21}, \cite{b22} minimum 100 stars \cite{b23}, \cite{b24}, minimum 100 commits \cite{b25}, minimum 5 contributors \cite{b26}, and minimum 2 branches \cite{b27}.

\begin{comment}
To focus on projects with active, sustained development and meaningful collaboration, we used a filtering process based on established practices for open-source repository curation. We applied the following five criteria sequentially:
\begin{itemize}
  \setlength\itemsep{0pt} % remove extra space between items
  \setlength\itemindent{1em} % small horizontal indent
  \item Recent commit activity (2024 or later): To focus on currently maintained projects, as inactive repositories do not reflect current development practices and ML evolves quickly \cite{b21}.
  \item Minimum 100 stars: While stars alone do not indicate quality, they show community recognition and usage, with prior research linking higher popularity to better maintenance and documentation \cite{b22},\cite{b23}.
  \item Minimum 100 commits: To ensure development history beyond quick experiments, as repositories with very few commits are often incomplete \cite{b24}.
  \item Minimum 5 contributors: Projects with fewer than 3-5 developers typically lack the team dynamics necessary for collaboration studies \cite{b25}. We kept this threshold moderate since many ML projects operate with smaller teams \cite{b26}.
  \item Minimum 2 branches: Multiple branches suggest organized workflows such as feature branches or release management, indicating a well-maintained codebase \cite{b27}.
\end{itemize}
\end{comment}
These criteria reduced the dataset from 572 to 279 repositories, resulting in a collection of active, collaborative ML projects suitable for code quality analysis. The filtered dataset displays diverse characteristics, as shown in Table~\ref{tab:dataset_characteristics}. The wide range across all metrics indicates the successful capture of both smaller collaborative projects and large-scale community efforts. 
\begin{table}[htbp]
\caption{Filtered Dataset Characteristics}
\begin{center}
\begin{tabular}{|c|c|c|c|}
\hline
\textbf{Metric} & \textbf{Min} & \textbf{Median} & \textbf{Max} \\
\hline
Stars & 167 & 2,657 & 209,443 \\
\hline
Commits & 120 & 1,467 & 188,754 \\
\hline
Branches & 2 & 19 & 1,766 \\
\hline
Contributors & 5 & 53 & 3,231 \\
\hline
LOC & 55 & 16,576 & 2,109,291 \\
\hline
\end{tabular}
\label{tab:dataset_characteristics}
\end{center}
\end{table}

To enable comparative analysis of code smells across different project scales, we categorized repositories by lines of code using percentile-based thresholds. The 30th and 60th percentiles divided the dataset into three balanced-sized groups (Table~\ref{tab:project_size}). 

\begin{table}[htbp]
\caption{Project Size Distribution by LOC}
\begin{center}
\begin{tabular}{|c|c|c|c|}
\hline
\textbf{Size Category} & \textbf{LOC Range} & \textbf{Count} & \textbf{Percentage} \\
\hline
Small & $<$ 10,445 & 82 & 29.4\% \\
\hline
Medium & 10,445 -- 26,821 & 85 & 30.5\% \\
\hline
Large & $>$ 26,821 & 112 & 40.1\% \\
\hline
\end{tabular}
\label{tab:project_size}
\end{center}
\end{table}

Beyond the updated metrics and filtering criteria, we enriched the dataset with two additional features essential for analyzing code smell patterns across project contexts. First, we calculated project age as the time span between the repository creation date and the last commit date, enabling examination of how code quality evolves with project maturity. The project ages range from 5.36 to 16.40 years across all repositories, with median ages of 7.74 years for small projects, 6.89 years for medium-sized projects, and 7.42 years for large projects.

Second, we systematically updated the CI/CD classification from NICHE, which was outdated (collected in 2020, before widespread adoption of GitHub Actions) and lacked the binary categorization needed for statistical analysis. Following established CI/CD definitions \cite{b28},\cite{b29}, we focused on build, test, and deployment automation. One author developed an automated detection script validated through manual inspection of 30 projects (93.3\% accuracy), with ambiguous cases resolved by two authors. The final classification includes 244 projects (87.5\%) with CI/CD and 35 (12.5\%) without. Complete methodology and validation results are provided in our replication package.

Third, we categorized all 279 repositories into application domains to identify domain-specific code smell patterns. Since GitHub lacks standardized domain labels and over half of the repositories do not have topic tags, we adopted a two-step hybrid classification approach. First, we defined seven domains based on recognized ML categories and assigned relevant keywords to each (e.g., Computer Vision: yolo, mask-rcnn, opencv, segmentation; NLP: transformer, bert, gpt, sentiment). Using the GitHub REST API, we retrieved repository metadata such as descriptions, README content, and topic tags, then applied automated keyword-based matching \cite{b30},\cite{b31}. Projects were classified through priority indicators (high-confidence matches) and a weighted scoring system. This automated process classified 226 repositories (81.1\%). The remaining 53 repositories (18.9\%) were manually reviewed by two authors, who examined README files and resolved ambiguous cases through discussion to ensure comprehensive coverage. The final categories include seven domains: ML Frameworks/Libraries (95, 34.1\%), NLP (39, 14.0\%), Computer Vision (37, 13.3\%), Domain-Specific Applications (33, 11.8\%), Data Science/Traditional ML (27, 9.7\%), MLOps/Deployment (26, 9.3\%), and Reinforcement Learning (RL) (22, 7.9\%). The Domain-Specific Applications group encompasses ML projects designed for specialized fields, including bioinformatics, robotics, signal processing, IoT/home automation, and healthcare, each with too few samples for separate analysis. Details of our classification process, domain keywords, and validation results are available in our replication package.

%%Table 3 presents the distribution of domains across different project sizes. ML Frameworks dominate the dataset, reflecting the infrastructure-centered focus of the ML open-source community.
\begin{comment}
\begin{table}[htbp]
\caption{Cross-Tabulation of Size Category × Domain}
\begin{center}
\begin{tabular}{|c|c|c|c|c|c|c|c|c|}
\hline
\textbf{Size} & \textbf{Frmwk} & \textbf{NLP} & \textbf{CV} & \textbf{Dom-Sp} & \textbf{DS} & \textbf{MLOps} & \textbf{RL} & \textbf{Total} \\
\hline
Small & 26 & 14 & 10 & 15 & 10 & 3 & 4 & 84 \\
\hline
Medium & 30 & 8 & 15 & 8 & 7 & 7 & 10 & 85 \\
\hline
Large & 39 & 17 & 12 & 10 & 10 & 16 & 8 & 112 \\
\hline
\textbf{Total} & 95 & 39 & 37 & 33 & 27 & 26 & 22 & 279 \\
\hline
\end{tabular}
\label{tab:size_domain}
\end{center}
\end{table}
\end{comment}

\subsection{Code Smell Detection}
To detect ML-CSs, we selected CodeSmile \cite{b12}, which identifies 12 of the 22 ML-CSs (Table~\ref{tab:ml_code_smells}) from Zhang et al.'s catalog \cite{b11}. Although MLpylint \cite{b19} detects more smell types (20 of 22) with higher precision (91.03\%), its very low recall (36.57\%) would miss many smell instances. SpecDetect4AI [19], which covers all 22 smells with better performance, was unavailable during our study. CodeSmile provides the best available balance of precision (82.13\%) and recall (75.79\%), making it suitable for large-scale analysis. The current GitHub implementation of CodeSmile extends the tool to detect 16 smell types; to maintain consistency with the published tool \cite{b12} and ensure comparability with prior work, we limited our analysis to the 
original 12 smell types validated in the published study (Table~\ref{tab:ml_code_smells}).

For assessing general Python code quality, we chose Pylint \cite{b8},\cite{b16}. Consistent with widespread static analysis practices \cite{b8},\cite{b32}, we excluded three categories of Pylint messages: (1) import-related warnings (including import-error, unused-import, wrong-import-order), because Pylint cannot reliably verify imports without the necessary project dependencies installed; (2) style and whitespace issues (such as invalid-name, line-too-long, bad-indentation), since these concern coding style rather than functional correctness; and (3) docstring-related alerts (like missing-class-docstring, missing-function-docstring), because documentation completeness does not directly impact code correctness. This filtering eliminated 4,209,804 messages, allowing us to focus on functional and logic-related quality concerns. We identified the top 20 most frequent functional and logic-related code smells from Pylint (Table~\ref{tab:python_code_smells}). Detailed occurrence data for each smell type can be found in our replication package.

To ensure consistency across our analysis, we cloned all 279 repositories in November 2025. Both CodeSmile and Pylint were executed using their default configurations on all Python files within each repository. We collected total smell counts per project, per-file smell counts, smell density (smells per LOC), and smell type distributions for each tool. This comprehensive data collection enables both project-level and file-level analysis of code quality patterns.

\begin{table}[htbp]
\caption{ML-CSs Detected by CodeSmile}
\centering
\begin{tabularx}{\columnwidth}{|c|X|}
\hline
\textbf{Pipeline Stage} & \textbf{Code Smells} \\
\hline
Data Cleaning & Chain Indexing (CIDX), Columns and DataType Not Explicitly Set (CDTNE), Dataframe Conversion API Misused (DCA), In-Place APIs Misused (IPA), Matrix Multiplication API Misused (MMA), Merge API Parameter Not Explicitly Set (MAP), NaN Equivalence Comparison Misused (NAN), Unnecessary Iteration (UI) \\
\hline
Model Training & Gradients Not Cleared Before Backward Propagation (GNC), Memory Not Freed (MNF), Pytorch Call Method Misused (PC), TensorArray Not Used (TA) \\
\hline
\end{tabularx}
\label{tab:ml_code_smells}
\end{table}

\begin{table}[htbp]
\caption{Top 20 General Python Code Smells (Pylint)}
\centering
\begin{tabularx}{\columnwidth}{|c|X|}
\hline
\textbf{Category} & \textbf{Code Smells} \\
\hline
Error & no-member, unsubscriptable-object, syntax-error, no-value-for-parameter\\
\hline
Warning & unused-argument, redefined-outer-name, attribute-defined-outside-init, abstract-method, unused-variable, fixme, redefined-builtin, arguments-differ \\
\hline
Refactor & no-self-use, too-many-arguments, too-many-locals, too-few-public-methods, no-else-return, too-many-instance-attributes\\
\hline
Convention & protected-access, logging-fstring-interpolation\\
\hline
\end{tabularx}
\label{tab:python_code_smells}
\end{table}

\subsection{Data Analysis}
To address our four research questions, we analyzed code smell densities 
(normalized per 1,000 lines of code) and their relationships with project 
characteristics using correlation and group comparison tests. All statistical
tests were performed using Python 3.10 with SciPy version 1.15.3, with a 
significance level of $\alpha = 0.05$.

To address RQ1, we formulate the following hypothesis:
\noindent \textbf{H0.0:} There is no significant difference between ML-specific code smell density and general Python code smell density within the same projects.  

%\noindent \textbf{H1.0:} ML-specific code smell density and general Python code smell density differ significantly within the same projects.

To test this hypothesis, we use the Wilcoxon signed-rank test, a non-parametric paired test suitable for our data since each project has measurements for both smell types (ML-specific and Python), and the data do not follow a normal distribution. We assess the effect size with Cliff's Delta ($\delta$), interpreted as follows: negligible ($|\delta| < 0.15$), small ($0.15 \leq |\delta| < 0.33$), medium ($0.33 \leq |\delta| < 0.47$), and large ($|\delta| \geq 0.47$). For general Python smells, we focus on the top 20 most frequent Pylint warnings rather than all 200+ rules available, following van Oort et al.\cite{b8}, who showed that analyzing the most frequent smell types yields meaningful insights while maintaining a balanced comparison with the 12 ML-CSs detected by CodeSmile. This decision ensures we compare systematic, practically significant issues (12 vs.\ 20 smell types) instead of an unbalanced comparison (12 vs.\ 200+).

To address RQ2 and RQ3, we investigate how six project characteristics relate to smell density for both ML-specific and general Python smells. For continuous characteristics, we assess correlations using Spearman's $\rho$ and differences across predefined groups using Mann--Whitney U (two groups) or Kruskal--Wallis H (three or more groups) with Bonferroni-corrected post-hoc tests. For two-level group comparisons based on continuous variables (project age, number of contributors, and commit frequency), we categorize the groups into young versus mature, fewer versus more contributors, and low- versus high-activity groups by splitting the values at the median to ensure balanced group sizes. The median values for our dataset are: project age, 7.56 years; contributors, 53; and commit frequency, 15.78 commits per month. For categorical variables (CI/CD adoption, domain), we apply the same group tests. Effect sizes are reported using Cliff's Delta ($\delta$), interpreted consistently with RQ1: negligible ($|\delta| < 0.15$), small ($0.15 \leq |\delta| < 0.33$), medium ($0.33 \leq |\delta| < 0.47$), and large ($|\delta| \geq 0.47$). For domain-level analyses, we additionally report eta-squared ($\eta^2$) (small: $0.01 \leq \eta^2 < 0.06$; medium: $0.06 \leq \eta^2 < 0.14$; large: $\eta^2 \geq 0.14$) and apply chi-squared tests to examine associations between smell types and domains. All hypotheses are summarized in Table~\ref{tab:tests_summary}.

\setlength{\tabcolsep}{3pt} 

\begin{table}[htbp]
\caption{Null hypotheses for RQ2--RQ3}
\centering
\footnotesize
\begin{tabularx}{\columnwidth}{|c|X|}
\hline
\textbf{Hypothesis} & \textbf{Null Hypothesis} \\
\hline
H1.0 / H1.1 & No correlation between project size and smell density \\
\hline
H1.2 / H1.3 & No difference in smell density across Small/Medium/Large projects \\
\hline
H2.0 / H2.1 & No correlation between project age and smell density \\
\hline
H2.2 / H2.3 & No difference in smell density between Young and Mature projects \\
\hline
H3.0 / H3.1 & No correlation between number of contributors and smell density \\
\hline
H3.2 / H3.3 & No difference in smell density between projects with Fewer vs. More contributors \\
\hline
H4.0 / H4.1 & No correlation between commit frequency and smell density \\
\hline
H4.2 / H4.3 & No difference in smell density between Low vs. High activity projects \\
\hline
H5.0 / H5.1 & No difference in smell density based on CI/CD adoption (Yes vs. No) \\
\hline
H6.0 / H6.1 & No difference in smell density among 7 domains \\
\hline
H6.2 / H6.3 & No association between domain and smell type \\
\hline
\end{tabularx}
\label{tab:tests_summary}
\end{table}

To systematically compare how ML-specific versus general Python smells relate to project characteristics (RQ4), we formulate three hypotheses (H7-H9) that test for different patterns between the two smell types.

\noindent \textbf{H7.0 (Continuous Characteristics):} There is no statistically significant difference between the correlation of ML-specific smell density with project characteristic $X$ and the correlation of general Python smell density with characteristic $X$, where 
\[
X \in \left\{
\parbox{0.45\columnwidth}{project size, project age, number of contributors, commit frequency}
\right\}.
\]

\noindent \textbf {H8.0:} There is no statistically significant difference between the effect of CI/CD adoption on ML-specific smell density and its effect on general Python smell density.

\noindent \textbf {H9.0:} There is no statistically significant association between domain ranking for ML-specific smell density and domain ranking for general Python smell density.

For H7, we compare the strength of correlations using Fisher's $z$-transformation to determine if two independent Spearman correlation coefficients ($\rho_\text{ML}$ and $\rho_\text{Python}$) are significantly different. The effect size is measured as $\Delta \rho = |\rho_\text{ML} - \rho_\text{Python}|$, with thresholds: negligible ($\Delta \rho < 0.1$), small ($0.1 \leq \Delta \rho < 0.3$), moderate ($0.3 \leq \Delta \rho < 0.5$), and large ($\Delta \rho \geq 0.5$).  

For H8, we compare Cliff's Delta values from H5 using bootstrap confidence intervals with 1,000 iterations, calculating $\Delta \delta = |\delta_\text{ML} - \delta_\text{Python}|$. For H9, we calculate Spearman's $\rho$ between domain median rankings for both smell types, identify domains with discordant patterns, and perform Mann--Whitney U tests with Bonferroni correction for discordant domain pairs.

Following common practice in empirical software engineering \cite{b33}, we use non-parametric tests throughout our analysis. These tests are robust against violations of normality assumptions and are well-suited for software metrics data, which often contain outliers and skewed distributions. For multiple comparisons, we apply Bonferroni correction to control the family-wise error rate when conducting post-hoc pairwise tests following significant Kruskal--Wallis results (H1.2/H1.3 for size categories, H6.0/H6.1 for domains, H9 for discordant domain pairs). The Bonferroni-adjusted significance threshold is calculated as $\alpha / k$, where $k$ is the number of pairwise comparisons. We report effect sizes for all tests to distinguish between statistical significance and practical importance, following the interpretation guidelines specified for each test type.

\section{Results}
This section presents our findings, which address the four research questions.
\subsection {RQ1. How do the densities of ML-CSs and general Python code smells compare across ML projects?}
To evaluate ML-specific versus general Python quality issues, we analyzed smell densities across 279 projects using paired tests. Table~\ref{tab:median_density_size} shows median densities for both smells in different project sizes.

\begin{table}[htbp]
\caption{Median code smell density by project size}
\centering
\footnotesize
\begin{tabular}{|l|c|c|c|}
\hline
\textbf{Size Category} & \textbf{$n$} & \textbf{ML-Specific} & \textbf{General Python} \\
\hline
Small  & 82  & 0.92 & 37.38 \\
Medium & 85  & 0.64 & 42.82 \\
Large  & 112 & 0.39 & 36.48 \\
\hline
\end{tabular}
\label{tab:median_density_size}
\end{table}
The Wilcoxon signed-rank test shows a highly significant difference between the two smell types ($W = 0.0$, $p < 0.001$, Cliff's $\delta = -0.979$, large effect). H0.0 is rejected.

\textbf{Interpreting the patterns.} Three key observations emerge from this comparison. First, ML-CSs are much less frequent than general Python smells, appearing 41 to 94 times less often, depending on the project size. Second, both types of smells tend to decrease as projects grow larger, but their patterns differ: ML-CSs drop by 58\% from small to large projects (0.92 to 0.39 smells/KLOC), while general Python smells remain relatively constant with only a 2\% decline (37.38 to 36.48 smells/KLOC). Third, the difference between the two smell types increases with project size, with the ratio shifting from 1:41 in small projects to 1:94 in large projects. This consistent frequency gap suggests that ML projects mainly encounter traditional code quality issues rather than ML-CSs. However, this low frequency should not imply they are unimportant. As shown in RQ2, ML-specific issues have fundamentally different relationships with project features than general Python problems, indicating they stem from distinct development challenges and require specialized mitigation strategies.
\begin{tcolorbox} [colback=gray!10, colframe=black, title=Finding 1] 
ML-CSs occur 41--94 times less often than general Python smells, representing only 1--2\% of detected issues. The growing gap in larger codebases suggests that ML-specific quality practices scale better with project growth than general Python quality practices.
\end{tcolorbox}

\subsection{RQ2. What are the relationships between ML-CSs and project characteristics?}
To identify factors affecting ML-specific code quality, we analyzed correlations and group differences related to six project characteristics. Table~\ref{tab:stat_tests_summary} provides a summary of all the statistical tests.

\begin{table}[htbp]
\centering
\footnotesize
\begin{threeparttable}
\caption{Summary of Statistical Tests for Project Characteristics and ML-Specific Smell Density}
\label{tab:stat_tests_summary}
\begin{tabularx}{\columnwidth}{|l|X|X|}
\hline
\textbf{Characteristic} & \textbf{Correlation Test} & \textbf{Group Comparison Test} \\
\hline
Size (LOC) 
& $\rho=-0.071$, $p=0.239$ 
& $H=3.574$, $p=0.167$ \\
\hline
Age (years) 
& $\rho=-0.035$, $p=0.562$ 
& $U=10541.5$, $p=0.228$; $\delta=0.083$ \\
\hline
Contributors 
& $\rho=-0.107$, $p=0.074$ 
& $U=10690.5$, $p=0.153$; $\delta=0.099$ \\
\hline
Commit Frequency 
& $\rho=-0.155$, $p=0.010$\tnote{a} 
& $U=11169.5$, $p=0.032$\tnote{a}; $\delta=0.148$ \\
\hline
CI/CD Adoption 
& --- 
& $U=3745.0$, $p=0.239$; $\delta=-0.123$ \\
\hline
Domain 
& --- 
& $H=21.184$, $p=0.002$\tnote{c} \\
\hline
\end{tabularx}
\begin{tablenotes}
\footnotesize
\item[a] Statistically significant at $p<0.05$
\item[b] Statistically significant at $p<0.01$
\item[c] Statistically significant at $p<0.001$
\end{tablenotes}
\end{threeparttable}
\end{table}

\textbf{Strong relationships: development activity and domain.} Two characteristics are significantly associated with ML-specific smell density. Commit frequency had a notable negative correlation ($\rho=-0.155$, $p=0.010$) and differed significantly between low-activity and high-activity projects ($U=11169.5$, $p=0.032$, $\delta=0.148$). Projects with high activity ($>15.78$ commits/month) exhibited 48\% lower smell density than less active projects (0.403 vs.\ 0.778 smells/KLOC), suggesting an association between development activity and ML-specific smell density, though this may also reflect that more active maintainers tend to be more proactive to identifying and addressing code smells.

Differences across domains were also significant. The Kruskal-Wallis test indicated significant differences among seven ML application areas ($H=21.184$, $p=0.002$, $\eta^2=0.072$, medium effect size). Post-hoc pairwise analyses with Bonferroni correction ($\alpha=0.0024$) found two significant pairs, both involving Data Science/Traditional ML, which consistently had the highest smell density (median: 1.955 smells/KLOC, $n=27$), approximately 2.4 to 5.3 times higher than other domains. RL had the lowest density (median: 0.370  smells/KLOC, $n=22$), followed by ML Frameworks (0.379, $n=95$) and NLP (0.487, $n=39$). We reject H4.0 and H4.2 (commit frequency), as well as H6.0 and H6.2 (domain).
\begin{figure*}[t]
  \centering
  \includegraphics[width=0.75\textwidth]{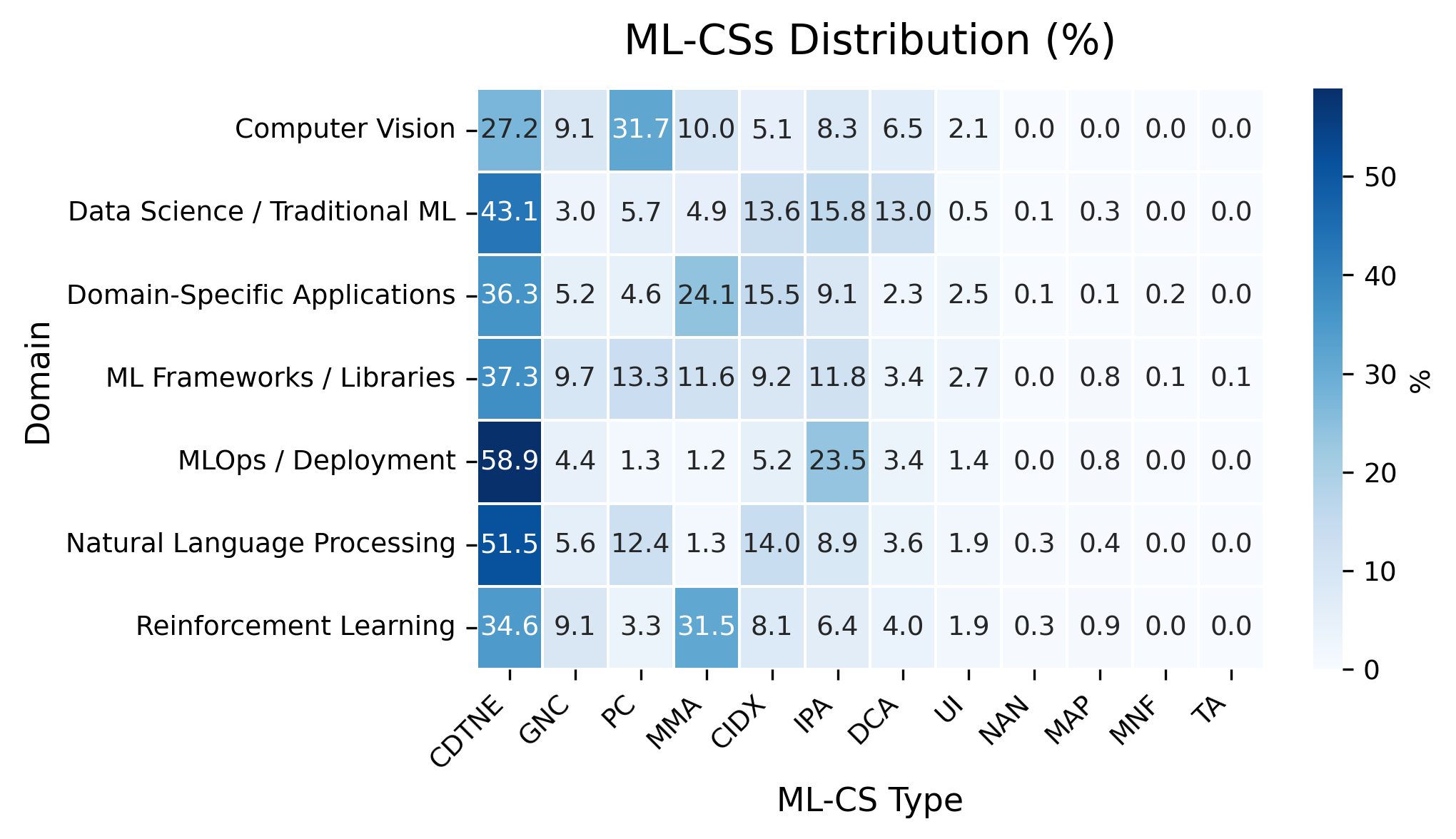} 
  \vspace{-3mm}
  \caption{Distribution of ML-CSs Across Domains.}
  \vspace{-4mm}
  \label{fig:fig2}
\end{figure*}
The chi-square test verified that smell type distributions differ significantly across various domains ($\chi^2=3802.34$, $p<0.001$, Cramer's $V=0.200$, indicating a moderate association). Figure~\ref{fig:fig2} reveals a striking pattern: ``Columns and DataType Not Explicitly Set'' dominates six of seven domains (34.6--58.9\%), with peaks in MLOps (58.9\%) and NLP (51.5\%), indicating that explicit schema management is a pervasive gap in ML development. Computer Vision is the exception, where GPU resource issues dominate (``Pytorch Call Method Misused'': 31.7\%). Secondary patterns show domain-specific concerns: RL balances configuration (34.6\%) with tensor manipulation (``Matrix Multiplication API Misused'': 31.5\%), MLOps exhibits data processing challenges (``In-Place APIs Misused'': 23.5\%), and Data Science/Traditional ML shows varied patterns consistent with exploratory workflows. Complete smell distributions are in our replication package.
\textbf{No significant relationships: Size, contributors, age, and CI/CD adoption.} Four characteristics did not relate to ML-specific smell density. Project size in lines of code showed a negligible correlation ($\rho=-0.071$, $p=0.239$), and size category comparisons found no significant differences ($H=3.574$, $p=0.167$). Similarly, contributor count exhibited a weak correlation ($\rho=-0.107$, $p=0.074$) with no significant group differences ($U=10690.5$, $p=0.153$, $\delta=0.099$). These null findings suggest that neither project scale nor team size reliably predicts ML-specific code quality. Project age had almost no correlation ($\rho=-0.035$, $p=0.562$) and showed no significant difference between young and mature projects ($U=10541.5$, $p=0.228$, $\delta=0.083$). This challenges traditional views of technical debt, which suggest quality usually declines over time. Likewise, CI/CD adoption had no effect ($U=3745.0$, $p=0.239$, $\delta=-0.123$), and projects with CI/CD (median: 0.532 smells/KLOC) had comparable quality to those without (median: 0.737 smells/KLOC). We fail to reject H1.0, H1.2 (size), H2.0, H2.2 (age), H3.0, H3.2 (contributors), and H5.0 (CI/CD).\\
\textbf{Interpreting the patterns.} The significant association of commit frequency, contrasted with the 
negligible effect of project age, suggests that continuous development activity is associated with higher code quality rather than project maturity alone. This aligns with the fast-changing ML ecosystem, where active projects regularly update frameworks and adopt new best practices. Inactive projects, however, tend to accumulate technical debt regardless of their age, which deviates from conventional technical debt theory \cite{b34}. The CI/CD null result, despite its documented benefits \cite{b28},\cite{b29}, indicates that standard pipelines confirm integration but do not ensure ML-specific correctness, such as tensor shapes, memory cleanup, or determinism. Furthermore, the lack of findings regarding project size and team size indicates that scale alone does not dictate ML-specific quality; rather, development activity and domain-specific practices are what truly matter. Data Science/Traditional ML projects show 2.4--5.3$\times$ higher smell density than specialized domains (1.955 vs.\ 0.370--0.831 smells/KLOC), implying that exploratory workflows build up more technical debt compared to production-focused practices.
\vspace{-4mm}
\begin{tcolorbox}[colback=gray!10, colframe=black, title=Finding 2]
High-activity projects exhibit significantly lower smell density, while age has no impact, contradicting traditional technical debt theory that suggests quality declines over time.
\end{tcolorbox}
\vspace{-2mm}
\begin{tcolorbox}[colback=gray!10, colframe=black, title=Finding 3]
Standard CI/CD pipelines seem not to reduce ML-specific technical debt. Despite benefits for traditional software quality, CI/CD adoption has no effect, revealing a testing gap in which pipelines validate integration but not ML-specific correctness.
\end{tcolorbox}
\begin{tcolorbox}[colback=gray!10, colframe=black, title=Finding 4]
Domains exhibit distinct smell patterns, with Data Science/Traditional ML showing higher density than specialized fields. While configuration issues are prevalent across most domains, smell profiles highlight specific challenges: MLOps struggles with data schema problems, Computer Vision demands GPU resource management, and RL involves balancing configuration with tensor manipulation, each requiring specialized quality strategies.
\end{tcolorbox}
\subsection{RQ3. What are the relationships between general Python code smells and the same project characteristics?}
We used the same statistical analysis framework to study general Python code smells and their correlations with the six project characteristics examined in RQ2. Table~\ref{tab:python_smells_summary} provides a summary of all the results.

\textbf{Pervasive quality issues across all contexts.} Unlike ML-CSs discussed in RQ2, general Python smell density showed no significant correlation with any project attribute. All correlation coefficients were close to zero: size ($\rho=0.008$, $p=0.890$), age ($\rho=0.056$, $p=0.352$), contributors ($\rho=-0.026$, $p=0.672$), and commit frequency ($\rho=-0.024$, $p=0.684$). Group comparisons also found no significant differences for any characteristic. Projects with CI/CD (median: 37.45 smells/KLOC) had similar smell densities to those without (median: 42.33 smells/KLOC, $U=3719.0$, $p=0.217$). Median densities remained consistently high (between 35.4 and 42.3 smells/KLOC), regardless of project size, age, team makeup, or development activity. We fail to reject any null hypothesis (H1.1, H1.3, H2.1, H2.3, H3.1, H3.3, H4.1, H4.3, H5.1).

\begin{table}[htbp]
\centering
\footnotesize
\caption{Summary of Statistical Tests for General Python Code Smell Density}
\begin{tabularx}{\columnwidth}{|l|X|X|}
\hline
\textbf{Characteristic} & \textbf{Correlation Test} & \textbf{Group Comparison Test} \\
\hline
Size (LOC) & $\rho=0.008$, $p=0.890$ & $H=4.328$, $p=0.115$ \\
\hline
Age (years) & $\rho=0.056$, $p=0.352$ & $U=9178.0$, $p=0.413$; $\delta=-0.057$ (Negligible) \\
\hline
Contributors & $\rho=-0.026$, $p=0.672$ & $U=10290.0$, $p=0.406$; $\delta=0.058$ (Negligible) \\
\hline
Commit Frequency & $\rho=-0.024$, $p=0.684$ & $U=9940.0$, $p=0.756$; $\delta=0.022$ (Negligible) \\
\hline
CI/CD Adoption & --- & $U=3719.0$, $p=0.217$; $\delta=-0.129$ (Negligible) \\
\hline
Domain & --- & $H=5.314$, $p=0.504$ \\
\hline
\end{tabularx}
\label{tab:python_smells_summary}
\end{table}

\begin{figure*}[t]
  \centering
  \includegraphics[width=0.75\textwidth]{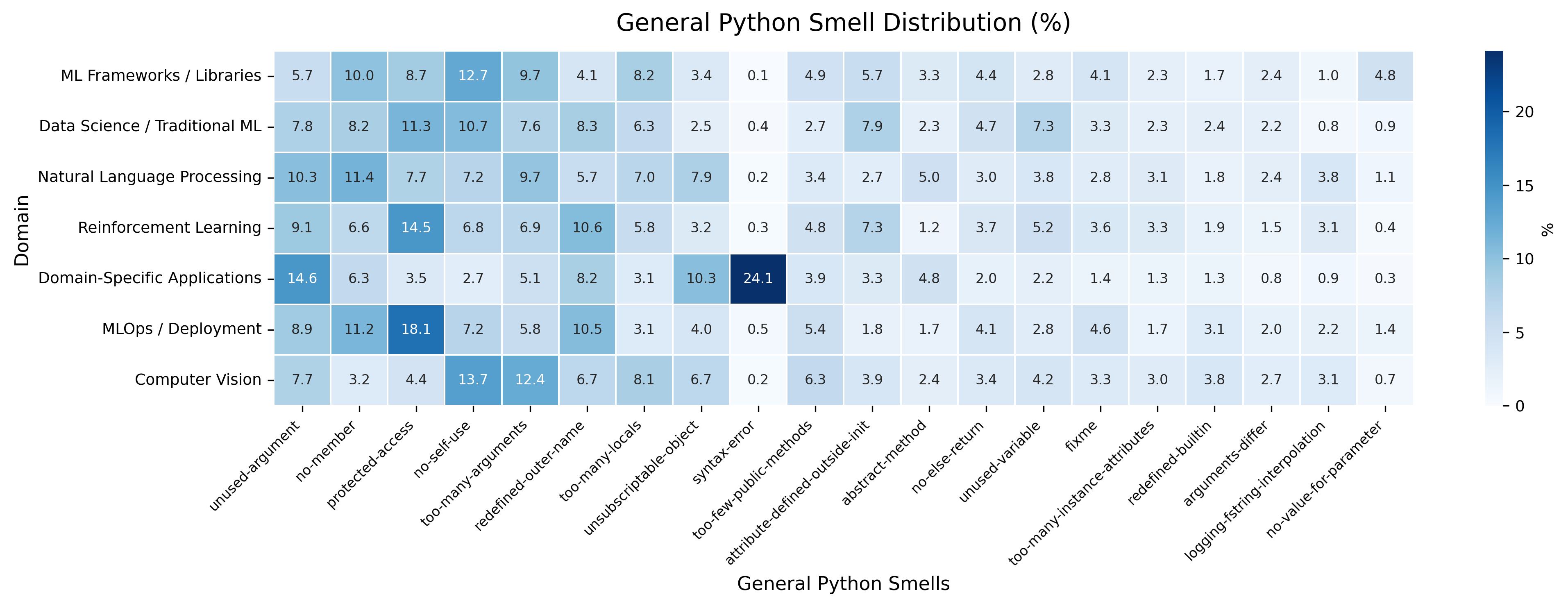}  % Was 0.93, now 0.75
  \vspace{-3mm}
  \caption{Distribution of General Python Smells Across Domains.}
  \vspace{-4mm}
  \label{fig:fig3}
\end{figure*}

The domain showed marginal differences ($H=5.314$, $p=0.504$) that did not reach statistical significance at $\alpha=0.05$. Median smell densities ranged from 35.38 smells/KLOC (ML Frameworks / Libraries) to 42.33 smells/KLOC (Computer Vision). Figure~\ref{fig:fig3} displays the distribution of smell types across domains, highlighting some domain-specific trends: Domain-Specific Applications shows the highest concentration with syntax errors (24.1\%), MLOps features concentrated protected-access issues (18.1\%), and RL shows increased issues with protected access (14.5\%). Data Science shows more evenly spread patterns related to complexity smells. However, unlike ML-CSs where domain effects were highly significant, these Python smell differences are relatively modest. We fail to reject H6.1.

\textbf{Interpreting the patterns.} The absence of notable relationships suggests that widespread Python quality problems are common across all types of ML projects. Unlike ML-CSs, which are influenced by development activity and vary by domain (RQ2), general Python smells are consistently problematic regardless of project specifics. This indicates fundamental differences in their origins: ML-CSs stem from domain-specific implementation issues and can be addressed through active maintenance, whereas general Python smells arise from common coding practices and complexity challenges that affect all project types. The high density of these smells (35--42 smells/KLOC, versus 0.4--2.0 for ML-specific issues) underscores that these are systemic concerns requiring distinct mitigation approaches from those for ML-specific quality problems.

\begin{tcolorbox}[colback=gray!10, colframe=black, title=Finding 5]
General Python code smells are widespread quality problems that are not influenced by the specific features of a project. Unlike ML-CSs, which fluctuate with development activity and vary by domain, Python smells tend to stay consistently high across all projects. This suggests underlying systemic issues in coding practices and complexity that need universal solutions instead of targeted fixes.
\end{tcolorbox}
\subsection{RQ4. What are the differences in how ML-CSs and general Python code smells correlate with project characteristics?}
We conducted formal tests to determine if project characteristics influence the two smell types differently, using Fisher's $z$-transformation (see Table~\ref{tab:differential_tests}). All comparisons indicated no significant differences. Therefore, we fail to reject hypotheses H7.0--H11.0.

\begin{table}[htbp]
\centering
\footnotesize
\begin{threeparttable}
\caption{Differential Effect Comparison Tests for ML-Specific and General Python Smells}
\label{tab:differential_tests} % <-- label right after caption
\begin{tabularx}{\columnwidth}{|l|X|X|X|}
\hline
\textbf{Characteristic} & \textbf{$\rho_\mathrm{ML}$} & \textbf{$\rho_\mathrm{Python}$} & \textbf{p-value} \\
\hline
Size & -0.071 & 0.008 & 0.353 \\
Age & -0.035 & 0.056 & 0.286 \\
Contributors & -0.107 & -0.026 & 0.334 \\
Commit Frequency & -0.155\tnote{a} & -0.024 & 0.122 \\
CI/CD ($\delta$) & -0.123 & -0.130 & 0.482 \\
\hline
\end{tabularx}
\begin{tablenotes}
\footnotesize
\item[a] $p<0.01$ for individual correlation
\end{tablenotes}
\end{threeparttable}
\end{table}

\begin{table*}[t]
\centering
\footnotesize
\begin{threeparttable}
\caption{Domain-Specific Smell Pattern Comparison}
\label{tab:domain_ranks}
\begin{tabular}{|l|c|c|c|c|c|>{\raggedright\arraybackslash}p{3.5cm}|}
\hline
\textbf{Domain} & \textbf{ML-CS Density} & \textbf{ML-CS Rank} & \textbf{Python Smell Density} & \textbf{Python Smell Rank} & \textbf{Rank Diff} & \textbf{Pattern} \\
\hline
Data Science/Traditional ML & 1.955 & 1 & 40.901 & 4 & 3 & ML-focus needed \\
Domain-Specific Applications & 0.831 & 2 & 41.881 & 2 & 0 & Both problematic \\
Computer Vision & 0.661 & 3 & 42.325 & 1 & 2 & Python-focus needed \\
MLOps/Deployment & 0.492 & 4 & 41.794 & 3 & 1 & Balanced \\
NLP & 0.487 & 5 & 37.647 & 6 & 1 & Python-focus needed \\
ML Frameworks/Libraries & 0.379 & 6 & 35.380 & 7 & 1 & Both well-managed \\
RL & 0.370 & 7 & 38.748 & 5 & 2 & Python-focus needed \\
\hline
\end{tabular}
\begin{tablenotes}
\footnotesize
\item Note: ML-CSs density and Python smell density are in smells/KLOC. Ranks are from highest (1) to lowest (7) density. Spearman's $\rho=0.643$, $p=0.119$ indicates independent patterns.
\end{tablenotes}
\end{threeparttable}
\end{table*}

Domain rankings indicated independence ($\rho=0.643$, $p=0.119$). Table~\ref{tab:domain_ranks} shows that domains problematic for ML-CSs are not always problematic for Python smells. For example, Data Science/Traditional ML ranks 1\textsuperscript{st} for ML-CSs but 4\textsuperscript{th} for Python smells, a difference of three ranks, while Computer Vision exhibits the opposite trend (3\textsuperscript{rd} for ML-CSs, 1\textsuperscript{st} for Python smells). We fail to reject H12.0, which confirms that their quality patterns are independent.

\textbf{Interpreting the patterns.} Domain ranking independence shows that the factors influencing ML-specific quality are different from those impacting general Python qualities. Some domains may perform well in preventing ML-specific problems but face challenges with traditional code quality, or the opposite. This implies these are separate quality aspects that need individual focus. 

\begin{tcolorbox}[colback=gray!10, colframe=black, title=Finding 6]
Domain quality patterns are distinct across different smell types, necessitating tailored strategies for each quality aspect.
\end{tcolorbox}
\section{Discussion}
Drawing from our review of 279 ML projects, we offer practical advice for practitioners aiming to enhance ML code quality and highlight important research opportunities for the software engineering community.
\subsection{Implications for Practitioners}
Our findings provide insights that may inform ML code quality practices:
\begin{itemize}
    \item \textbf{Implement dual-layer quality strategies.} Since ML-CSs form only 1--2\% of issues but cause silent failures, and general Python smells are common maintainability problems, organizations need two complementary approaches: (1) automated linters like Pylint deployed at scale through CI/CD for broad Python issues, and (2) specialized ML detection tools (e.g., CodeSmile or similar) integrated as pre-commit hooks or mandatory code review checks to catch rare but critical correctness problems.
    \item \textbf{Prioritize continuous maintenance over one-time quality initiatives. } Since commit frequency relates to ML-specific quality but project age does not, organizations can implement continuous practices such as integrating ML-specific linters into daily workflows, setting smell density thresholds that trigger code reviews, monitoring trends with monthly or quarterly KPIs, and scheduling refactoring whenever framework dependencies are updated.
    \item \textbf{Extend CI/CD pipelines with ML-specific quality gates.} Standard CI/CD adoption does not influence ML-specific smell density since typical pipelines mainly verify integration correctness and do not target properties unique to ML. Organizations can enhance their pipelines by: (1) making ML-specific linters mandatory as pre-merge checks, (2) implementing data schema validation with tools like Great Expectations to identify configuration issues, (3) validating data preprocessing steps to prevent chain indexing problems and unnecessary iterations, (4) monitoring GPU memory consumption with tools such as PyTorch Profiler or TensorFlow Memory Profiler to identify leaks, and (5) verifying tensor shapes and dimensions at model input and output stages. These recommendations align with established ML testing frameworks \cite{b35},\cite{b36} and MLOps best practices \cite{b37}, highlighting the importance of ML-specific validation beyond conventional software testing.
    \item \textbf{Adopt domain-tailored quality assurance strategies.} The domain significantly influences ML-specific quality. Because domain rankings are separate from general Python quality rankings, teams should: (1) benchmark their quality metrics against projects within the same domain rather than relying on overall ML averages (e.g., Data Science/Traditional ML projects exhibit higher baseline smell density due to notebook-driven workflows \cite{b38},\cite{b39}, requiring different targets than production-focused domains such as RL or Computer Vision), (2) conduct domain-specific code reviews, including schema validation for MLOps, tensor shape checks for RL, and GPU memory profiling for Computer Vision, and (3) develop testing strategies customized to each domain’s typical failure patterns.
\end{itemize}
\subsection{Implications for Researchers}
Our study opens several research directions:
\begin{itemize}
    \item \textbf{Improve integration of ML-specific detection tools into development workflows.} Despite tools like CodeSmile, MLpylint, and SpecDetect4AI \cite{b12},\cite{b19},\cite{b20}, their use is limited. Future research should integrate these by embedding ML-specific linters into pre-commit hooks and CI/CD pipelines and develop context-aware feedback systems to help developers detect ML-related anti-patterns during code reviews.
    \item \textbf{Investigate domain-specific best practices and their transferability.} Our discovery that specialized areas like RL and ML Frameworks have 4--5 times lower smell density than Data Science and Traditional ML prompts important questions: what practices lead to higher quality? Can notebook workflows adopt production methods without losing their exploratory nature? Qualitative studies on high-quality projects might reveal practices for wider adoption.
    \item \textbf{Replicate across programming languages and ML ecosystems.} Our Python-focused study doesn't explore if similar patterns exist in R, Julia, or MATLAB, how smell patterns differ across frameworks like PyTorch, TensorFlow, and JAX, or if domain-specific patterns stay consistent across ecosystems. Replication is needed to see if our findings reveal universal ML issues or are Python-specific, aiding in creating language-agnostic quality frameworks.
    \item \textbf{Investigate CI/CD pipeline quality in ML projects.} Our observation that the adoption of CI/CD does not correlate with ML-specific smell density indicates that simply adopting these practices is not enough. Future research should (1) examine the content and sophistication of ML CI/CD pipelines to identify practices that truly address ML-specific quality issues, and (2) create maturity models that differentiate between basic integration testing and comprehensive ML-aware quality assurance. These studies would help clarify which CI/CD practices most effectively enhance ML code quality.

\end{itemize}
\section{Threats to validity}
Our findings might be affected by the measurement tools used. CodeSmile detects 12 out of 22 ML-specific smell types, which could lead to an underestimation of their occurrence; however, no comprehensive alternative exists. To stay consistent with the original CodeSmile tool, we focused on the 12 smell types validated in the initial study, excluding four additional smells from the current GitHub version (Hyperparameters Not Explicitly Set, Empty Column Misinitialization, Broadcasting Feature Not Used, and 
Deterministic Algorithm Option Not Used). While this ensures 
comparability with prior work, their exclusion may affect the observed domain-specific patterns. For general Python quality, we used Pylint, a widely accepted tool in software engineering research. Although both tools may generate false positives or negatives, their extensive prior use supports their appropriateness for comparison. Classification accuracy involves additional risks. Domain classification combined automated methods (81.1\%) with manual validation (18.9\%), where two researchers resolved ambiguous cases by inspecting README files. CI/CD classification achieved 93.3\% validation accuracy based on literature criteria, with dual review for uncertain repositories. Some misclassification may still occur, but the systematic review process reduces subjective bias. The CI/CD dataset is considerably imbalanced, comprising 244 projects that use CI/CD and only 35 that do not. This imbalance might weaken the statistical power of analyses concerning CI/CD (H5.0, H5.1, H11.0) and make it more difficult to detect differences between groups. While systematic classification helps minimize labeling errors, the small count of non-CI/CD projects underscores how widely CI/CD is adopted in contemporary ML development.

The filtering criteria used for project selection (e.g., minimum 100 stars,100 commits, 5 contributors, and 2 branches) were based on established practices in the literature \cite{b21},\cite{b22},\cite{b23},\cite{b24},
\cite{b25},\cite{b26},\cite{b27} but were not subjected to sensitivity analysis. Different thresholds could result in a different set of projects and potentially affect the findings. We acknowledge this as a limitation and encourage future work to investigate the sensitivity of these results to varying selection criteria. Statistical validity may be influenced by test selection and sample size. Due to non-normal data distributions, we used non-parametric tests (Spearman, Mann–Whitney U, and Kruskal–Wallis H) and reported effect sizes, such as Cliff's Delta. A Bonferroni correction (setting $\alpha$ = 0.0024) was applied for multiple domain comparisons, which increased the risk of Type II errors. For RQ4, Fisher's z-transformation and bootstrap testing (1,000 iterations) were used. Smaller subgroups (e.g., RL: n = 22, MLOps: n = 26) may limit statistical power. External validity is limited by the sample choice. We examined 279 collaborative, actively maintained open-source GitHub projects across seven ML domains, varying in size, maturity, and team makeup, including major frameworks like PyTorch, TensorFlow, and scikit-learn. Nonetheless, the results may not extend to proprietary systems or ML projects in other programming languages.

%Despite these constraints, our study benefits from the dataset's scope, validated dual-review classifications, suitable statistical methods for non-normal data, transparent effect size reporting, and a publicly accessible replication package.
\section{Conclusion and Future Work}
We conducted an empirical study on 279 open-source machine learning projects to investigate the relationship between ML-specific and general Python code quality and various project characteristics. Although ML-CSs are less common than general Python smells, they present higher risks due to silent correctness errors and need specialized detection methods. Development activity is linked to ML code quality, whereas project age and team size are not, challenging conventional views on technical debt. We found no correlation between CI/CD adoption and ML-specific smell levels, highlighting a testing gap where standard pipelines miss domain-specific issues. Code quality also differs by domain: data science and traditional ML show higher smell density, while specialized areas have unique patterns, such as configuration in MLOps, tensor operations in RL, and GPU workflows in Computer Vision. The rankings of ML-specific and general Python smells are independent across domains, emphasizing the importance of domain-specific quality approaches. 

Future research can expand on this by using more advanced detectors like SpecDetect4AI \cite{b20} when available, validating findings on proprietary codebases, exploring other ML languages and frameworks, and creating ML-specific CI/CD quality gates based on domain smell patterns.

%\section*{References}

\end{document}